\documentclass[12pt]{iopart}

\usepackage[dvips]{graphics}

\begin{document}

\title[Chaotic Hamiltonian transport]{Non-Gaussian features of chaotic Hamiltonian transport\footnote{This work is dedicated to the memory of Prof. Radu Balescu.}}

\author{Roberto Venegeroles  and
Alberto Saa\footnote{On leave of absence from UNICAMP, Campinas, SP, Brazil.}}
\address{Centro de Matem\'atica, Computa\c c\~ao e Cogni\c c\~ao,
Universidade Federal do ABC, 09210-170, Santo Andr\'e, SP, Brazil}
\eads{\mailto{roberto.venegeroles@ufabc.edu.br},  \mailto{asaa@ime.unicamp.br}}

\begin{abstract}
Some non-Gaussian aspects of  chaotic transport are investigated for a general class of two-dimensional area-preserving maps. Kurtosis, in particular, is calculated from the diffusion and the Burnett coefficients, which are obtained  analytically. A characteristic time scale delimiting the onset of the
 Markovian   regime for the master equation  is established. Some explicit examples are discussed.
\end{abstract}

\pacs{05.45.Ac, 05.60.Cd, 05.20.-y}
\maketitle

\section{Introduction}

Discrete-time systems  have a  prominent role in many branches
of  nonlinear sciences. Hamiltonian (or area-preserving) maps,
for instance,
are particularly relevant for the  modeling of classical dynamical systems\cite{Lichtenberg,Reichl}.
The study of chaos in such systems has followed two main lines.
The first one considers  some individual trajectories   in order to explore and characterize the system main topological properties\cite{Lichtenberg,Reichl}. The
second one investigates the distribution functions of   statistical ensembles or, more specifically, some transport properties of the associated maps\cite{Gaspard}.
The present work belongs to this latter group.

In the past,
many  investigations of   transport properties for Hamiltonian maps have been motivated
 by the paradigmatic Chirikov-Taylor standard map\cite{Chirikov}. Although considerable progress in the study of
 diffusion has been achieved in the last 25 years \cite{Rechester,Abarbanel,Cary,McKay,Hatori,AAC,Dana,KD},
  higher order transport   coefficients have been particularly
  overlooked. The the fourth order coefficient $B$ known as the Burnett coefficient\cite{Gaspard}, for instance, plays a central role in the large deviations theory: its magnitude gives the first indication of the deviation of a density function from a Gaussian packet.

  Here, we consider some non-Gaussian features of the chaotic transport for Hamiltonian maps by means
  of high order corrections to the spectral properties of the associate  Perron-Frobenius operator $U$.
  Exponential relaxation for $U$ was rigorously established for hyperbolic systems by Pollicott and Ruelle \cite{Pollicott,Ruelle}. The relaxation rates $\gamma_{m}$, known as Pollicott-Ruelle resonances, are related to the poles   $z_{m}=e^{\,\gamma_{m}}$ of   the resolvent $R(z)=(z-U)^{-1}$. These resonances are located inside the unit circle in the complex $z$ plane, despite that the spectrum of $U$ is confined to the unit circle because of unitarity\cite{Hasegawa2}.
  The normal late time evolution of density or correlations functions are dominated by the leading Pollicott Ruelle resonances (LPR).
    Recently, this mathematically well-established results has been confirmed in the high stochasticity approximation for some mixed
  systems\cite{Khodas,Venegeroles}. In    \cite{Venegeroles}, the
LPR resonances are
  analytically  calculated
for the generic
  radial twist map\cite{Lichtenberg}
\begin{eqnarray}
\begin{array}{l} I_{n+1} = I_{n}+K\,f(\theta_{n})\,,\\
                 \theta_{n+1} = \theta_{n}+c\,\alpha(I_{n+1})\qquad \mbox{mod}\, 2\pi,
\end{array} \label{map}
\end{eqnarray}
defined on the cylinder $-\pi\leq\theta<\pi$, $-\infty<I<\infty$.
We call  $f(\theta)$  and $\alpha(I)=\alpha(I+2\pi r)$, respectively,
  the impulse function and the rotation number. The constants
$c$, $r$, and $K$  are assumed to be real, and $K$ is named   the stochasticity parameter.
One can also consider non-periodic cases  by taking  the limit $r\rightarrow\infty$.
 The LPR resonances for (\ref{map}) were obtained in \cite{Venegeroles} without any high stochasticity approximation,
 up to corrections of second order in the wavenumber. It is shown, in particular,
that the wavenumber dependence of the LPR  resonances determines the transport coefficients.

In the present paper, we extend the results of \cite{Venegeroles} to
 higher order wavenumber corrections  with the purpose of  evaluating   the
 Burnett coefficient  for the map (\ref{map}). Kurtosis, in particular, is then
 explicitly calculated.
 Our results are compared with   numerical
 simulations for some specific models, namely the standard map, the sawtooth map,
 and two maps with non-linear rotation numbers: a periodic one (the tangent map), and
   a non-periodic one (the cubic map). In all cases, a very good agreement is obtained.
  Our results allow us also to infer  a characteristic time scale  delimiting  the onset
  of the Markovian regime for the density function. We show that such a characteristic time scale
  is sharper than the others previously obtained  in the literature.

\section{Statistical Analysis}

The statistical analysis of the map (\ref{map}) is best carried out in Fourier space. 
The conditional probability   that an initial state $(I_{0},\theta_{0})$ evolves to a final state
$(I_n,\theta_n)$ is given by $\int dId\theta\,\rho_{n}(I,\theta)$. 
The Fourier expansion of the distribution function $\rho_n(I,\theta)$  can be written as
\begin{eqnarray}
\label{distribution}
\rho_{n}(I,\theta)=\sum_{m}\int dq\,e^{i(m\theta+qI)}a_{n}(m,q)\,,\\\nonumber
\end{eqnarray}
where the initial density is given by $\rho_{0}=\delta(I-I_{0})\delta(\theta-\theta_{0})$, and thus $a_{0}(m,q)=(2\pi)^{-2}e^{-i(m\theta_{0}+qI_{0})}$. The expected values of the moments $I^{p}$   can be calculated from the Fourier amplitudes $a_{n}(m,q)$ by
\begin{equation}
\label{moment}
\langle I^{\,p}\rangle_{n}=(2\pi)^{2}\,[(i\,\partial_{q})^{p}\,a_{n}(q)]_{q=0},
\end{equation}
where $a_{n}(q)\equiv a_{n}(0,q)$. An alternative way to calculate the moments is given by the following expression
\begin{equation}
\label{vanhove}
F_{n}(q)\equiv\left\langle\exp[-iq(I-I_{0})]\right\rangle_{n}\,,
\end{equation}
known as the Van Hove incoherent intermediate scattering function \cite{Gaspard}.
In the limit $q\rightarrow0$, $F_{n}(q)$ becomes the generating function of the generalized moments:
\begin{equation}
\label{generating}
F_{n}(q)=\exp\sum_{l=1}^{\infty}\frac{(-iq)^{l}}{l!}\,C_{l}(n),
\end{equation}
where $C_{l}$ denotes the cumulant moments\cite{McLennan,vanBeijeren}. Defining $\Delta I\equiv I-I_{0}$, the first cumulants are given by
\begin{eqnarray}
C_{1}&=&\left\langle\Delta I\right\rangle_{n},\\
C_{2}&=&\left\langle(\Delta I)^{2}\right\rangle_{n}-C_{1}^{2},\\
C_{3}&=&\left\langle(\Delta I)^{3}\right\rangle_{n}-3C_{1}C_{2}-C_{1}^{3},\\
C_{4}&=&\left\langle(\Delta I)^{4}\right\rangle_{n}-3C_{2}^{2}-4C_{1}C_{3}-6C_{1}^{2}C_{2}-C_{1}^{4}.
\end{eqnarray}
The existence of cumulant moments satisfying the equation (\ref{generating}) is not assumed a priori.
Note, however,
 that $F_{n}(q)$ is analytic around the origin $q=0$ if and only if all moments $\left\langle(\Delta I)^{l}\right\rangle_{n}$ exist and are finite. This condition breaks down in cases of distributions with ``fat tails", like the non-Gaussian Levy stable distributions\cite{Balescu1}. The Van Hove function (\ref{vanhove}) can be calculated explicitly from the density (\ref{distribution})
\begin{equation}
\label{Fnqanq}
F_{n}(q) = \int dId\theta\,e^{-iq(I-I_{0})}\rho_{n}(I,\theta)
 = (2\pi)^{2}e^{iqI_{0}}a_{n}(q).
\end{equation}
Assuming that the evolution law  for the relevant Fourier amplitude $a_{n}(q)$ is exponential for long times,
\begin{equation}
\label{anq}
a_{n}(q)=\exp[n\gamma(q)]a_{0}(q),
\end{equation}
 and that the initial relevant amplitude is given by $a_{0}(q)=(2\pi)^{-2}\,e^{-iqI_{0}}$,
 the dispersion rate $\gamma(q)$ can be obtained from the limit $n\rightarrow\infty$
 of the
 the Van Hove function
\begin{equation}
\label{gammaF}
\gamma(q)=\lim_{n\rightarrow\infty}\frac{1}{n}\ln[F_{n}(q)].
\end{equation}
Combining the generating function (\ref{generating}) with the dispersion rate (\ref{gammaF}),  one
 can define the generalized transport coefficients $\mathcal{D}_{2l}$ by
\begin{equation}
\label{disprel}
\mathcal{D}_{2l}\equiv\lim_{n\rightarrow\infty}\frac{1}{n}\frac{C_{2l}(n)}{(2l!)}=\frac{(-1)^{l}}{(2l)!}\,\partial_{q}^{2l}\gamma(q)|_{q=0}.
\end{equation}
If $C_{1}=0$, the Einstein formula for the diffusion coefficient $D$ is obtained by setting $l=1$,
\begin{equation}
\label{Ddef}
D \equiv \lim_{n\rightarrow\infty}\frac{1}{2n}\left\langle(\Delta I)^{2}\right\rangle_{n}=-\frac{1}{2}\partial_{q}^{2}\gamma(q)|_{q=0},
\end{equation}
while
 the Burnett coefficient $B$ is obtained for $l=2$,
\begin{equation}
\label{Bdef}
B \equiv \lim_{n\rightarrow\infty}\frac{1}{4!n}\left[\left\langle(\Delta I)^{4}\right\rangle_{n}-3\left\langle(\Delta I)^{2}\right\rangle_{n}^{2}\right]=\frac{1}{4!}\partial_{q}^{4}\gamma(q)|_{q=0}.
\end{equation}
The diffusion $D$ and the Burnett $B$ coefficients are the bases for our analysis
on non-Gaussian features of the chaotic transport for the map (\ref{map}).

\subsection{Accelerator modes}

Before starting our analysis, however,
one should warn about the so-called accelerator modes\cite{Lichtenberg},
corresponding to  fixed points $(\theta_{*},I_{*})$ of (\ref{map}):
\begin{equation}
Kf(\theta_{*})=2\pi rL_{I},\qquad c\,\alpha(I_{*})=2\pi L_{\theta},
\label{accelmodes}
\end{equation}
where $L_{\theta}$ and $L_{I}$ are integers satisfying the stability condition
\begin{equation}
|2+cKf'(\theta_{*})\alpha'(I_{*})|\leq2.
\label{stab}
\end{equation}
Typically, trajectories diffuse normally, although some of them may be dragged along the accelerated modes, if they do exist. These rare events   become meaningful for sufficiently high time scales, resulting in {  anomalous} diffusion of Levy type for map parameters satisfying (\ref{stab}).
In such a case, the diffusion coefficient behaves locally between normal dynamics and accelerator modes,
 for which one has $D\sim n^{\beta-1}$ for $1<\beta<2$ \cite{Balescu1,Zaslavsky}. In the case of the standard map, these divergences  result in peaks for the value of
$D$ for $K=2m\pi$, with decreasing
amplitude as $K$ increases\cite{Lichtenberg}.

\section{Higher order Pollicott-Ruelle Resonances}

The dispersion rate (\ref{gammaF}) for the system (\ref{map}) was considered in \cite{Venegeroles}
up to order $\mathcal{O}(q^{2})$
by means of the decomposition  of the resolvent $R(z)=(z-U)^{-1}$,
 based on the projection operator techniques utilized by Hasegawa and Saphir\cite{Hasegawa1} and
 Balescu\cite{Balescu1} for the standard map. The operator $U$ defines the law of evolution of the Fourier amplitudes, $a_{n}(q)=U^{n}a_{0}(q)$.
 Its iteration $U^{n}$ can be formally
 obtained through the   identity $\oint_{C} dz R(z)z^{n}=2\pi i U^{n}$, where the contour of integration lies outside the unit circle.
  One then introduces  the mutually orthogonal projection operators $P=\left|q,0\right\rangle\left\langle q,0\right|$, which selects the {  relevant} state, and its complement $Q=1-P$, leading to
   \begin{eqnarray}
a_{n}(q)=\frac{1}{2\pi i}\oint_{C}dz\,\frac{z^{n}}{z-\sum_{j=0}^{\infty}z^{-j}\Psi_{j}(q)}\,a_{0}(q),\label{relevantevolution}
\end{eqnarray}
where $\Psi_{j}(q)$ are the so-called memory functions for the system (\ref{map}).
  The resulting integral is  solved by the method of residues by truncating the infinite denominator series at $j=N$ and then taking the limit $N\rightarrow\infty$. The nontrivial leading pole was evaluated in the limit $n\rightarrow\infty$ by the well known Newton-Raphson iterative method starting
   with $z_{0}=1$ \cite{Venegeroles}.

 The $\mathcal{O}(q^{4})$ correction of the LPR resonances can be obtained by introducing
into the denominator of the equation (\ref{relevantevolution}) the $\mathcal{O}(q^{4})$ corrections
to the value of $z$,
\begin{equation}
z=1-Dq^{2}+\mathcal{O}(q^{4})
\end{equation}
 and repeating the same
steps done in \cite{Venegeroles}.
Taking into account
 that $\Psi_{0}(q)=1+\mathcal{O}(q^{2})$ and $\Psi_{j\geq1}(q)=\mathcal{O}(q^{2})$, the higher order LPR resonance  can be rewritten as
\begin{equation}
\label{gammaq6approx}
\gamma(q)=\ln\sum_{j=0}^{\infty}(1+jDq^{2})\Psi_{j}(q)+\mathcal{O}(q^{6}).
\end{equation}
The memory functions $\Psi_{j}(q)$ are the same ones obtained originally in \cite{Venegeroles} for the system (\ref{map})
 \numparts
\begin{equation}
\Psi_{0}(q)=\mathcal{J}_{0}(-Kq)\,,\label{psi0}
\end{equation}
\begin{equation}
\Psi_{1}(q)=\sum_{m}\mathcal{J}_{-m}(-Kq)\mathcal{J}_{m}(-Kq)\,\mathcal{G}_{0}(r,mc)\,,\label{psi1}
\end{equation}
\begin{eqnarray}
\label{psi2}
\Psi_{j\geq2}(q)&=&\sum_{\{m\}}\sum_{\{\lambda\}^{\dag}}\mathcal{J}_{-m_{1}}(-Kq)\,\mathcal{J}_{m_{j}}(-Kq)\,\mathcal{G}_{\lambda_{1}}(r,m_{1}c) \times \nonumber \\
&&\times\prod_{i=2}^{j}\mathcal{G}_{\lambda_{i}}(r,m_{i}c)\,\mathcal{J}_{m_{i-1}-m_{i}}\left[-K\left(q+r^{-1}\sum^{i-1}_{k=1}\lambda_{k}\right)\right]\label{psij}.
\end{eqnarray}
 \endnumparts
The Fourier decompositions of the $\alpha(I)$ and $f(\theta)$ functions are, respectively,
\begin{eqnarray}
\mathcal{G}_{l}(r,x)&=&\frac{1}{2\pi}\int d\theta\,\exp\{-i[x\alpha(r\theta)-l\theta]\}\,,\label{Gfunction}\\
\mathcal{J}_{m}(x)&=&\frac{1}{2\pi}\int\,d\theta\,\exp\{-i[m\theta-xf(\theta)]\}\,.\label{Jfunction}
\end{eqnarray}
Hereafter, the following convention is adopted: wavenumbers denoted by {   Roman indices} can only take {  non-zero} integer values, whereas wavenumbers denoted by {  Greek indices} can take {  all} integer values, including zero. For  fixed $j$, the sets of wavenumbers are defined by $\{m\}=\{m_{1},\ldots,m_{j}\}$ and $\{\lambda\}^{\dag}=\{\lambda_{1},\ldots,\lambda_{j}\}$, where the dagger   denotes the restriction $\sum_{i=1}^{j}\lambda_{i}=0$.
We introduce also the following series expansion for the function (\ref{Jfunction})
\begin{equation}
 \mathcal{J}_{m}(x)=\delta_{m,0}+\sum_{n=1}^{\infty}c_{m,n}\,x^{n},
 \end{equation}
 where
\begin{eqnarray}
c_{m,n}=\frac{1}{2\pi}\frac{i^{n}}{n!}\int\,d\theta\,f^{n}(\theta)\,e^{-im\theta}.
\label{coeficients}
\end{eqnarray}

\section{Transport Coefficients and Kurtosis}
\label{section3}

The general exact diffusion coefficient   can be obtained   by using
the definition (\ref{Ddef}) and the LPR resonance (\ref{gammaq6approx}),
 taking into account   (\ref{psi0})-(\ref{psi2}),
\begin{eqnarray}
\label{diffusionformula}
\frac{D}{D_{ql}}&=&1+2\sum_{m\geq1}\sigma_{m,m}\,\mbox{Re}[\mathcal{G}_{0}(r,mc)]+\sum^{\infty}_{j=2}\sum_{\{m\}}\sum_{\{\lambda\}^{\dag}}\,\sigma_{m_{1},m_{j}}\,\mathcal{G}_{\lambda_{1}}(r,m_{1}c)\times\nonumber \\
&&\times \prod_{i=2}^{j}\mathcal{G}_{\lambda_{i}}(r,m_{i}c)\,\mathcal{J}_{m_{i-1}-m_{i}}\left(-\frac{K}{r}\sum^{i-1}_{k=1}\lambda_{k}\right),
\end{eqnarray}
where the condition $c_{\,0,1}=0$ (and thus $C_{1}=0$) is  requeried, $D_{ql}=-c_{\,0,2}\,K^{2}$ is the quasilinear diffusion coefficient and $\sigma_{m,m'}=c_{-m,1} c_{m',1} /c_{0,2}$ \cite{Venegeroles}.
The diffusion coefficient (\ref{diffusionformula}) gives the lowest order macroscopic description of the diffusion process. If the evolution process is asymptotically truly diffusive, then the angle-averaged density should have a Gaussian contour after a sufficiently long time. A first indication of the deviation of a density function from a Gaussian packet is given by the Burnett coefficient $B$ defined by (\ref{Bdef}).
The dimensionless fourth order cumulant
\begin{equation}
\label{kurtosis}
\kappa(n)\equiv\frac{\left\langle(\Delta I)^{4}\right\rangle_{n}}{\left\langle(\Delta I)^{2}\right\rangle^{2}_{n}}
\end{equation}
is usually called the { kurtosis}. For Gaussian densities, $B=0$ and $\kappa=3$
for all times. Combining  (\ref{Ddef}), (\ref{Bdef}), and (\ref{kurtosis}), we obtain for sufficiently long times
\begin{equation}
\label{kurt}
\kappa=3+\frac{6B/D^{2}}{n}.
\end{equation}
The Burnett coefficient $B$ can be evaluated in the chaotic regime by truncating of the resonance $(\ref{gammaq6approx})$ at $j=2$ so that (\ref{Bdef}) can now be applied to (\ref{gammaq6approx}), yielding
\begin{equation}
\label{2trunc}
B\approx-\frac{5}{2}D^{2}+2DD_{ql}-\frac{1}{2}D\,\partial_{q}^{2}\Psi_{1}+
\frac{1}{4!}\partial_{q}^{4}(\Psi_{0}+\Psi_{1}+\Psi_{2})
\end{equation}
calculated at $q=0$.
 In order to verify these results,
 we   calculate the Burnett coefficients and its respective kurtosis for some particular cases of maps (\ref{map}) and compare with the respective numerical simulations. To this purpose, it is important
 to choose intermediate values of $n$, since exaggerated values   tends to wash non-Gaussian fluctuations
 away quickly.

\subsection{The standard map}

\begin{figure}[ht]
\begin{center}
\resizebox{0.65\linewidth}{!}{\includegraphics*{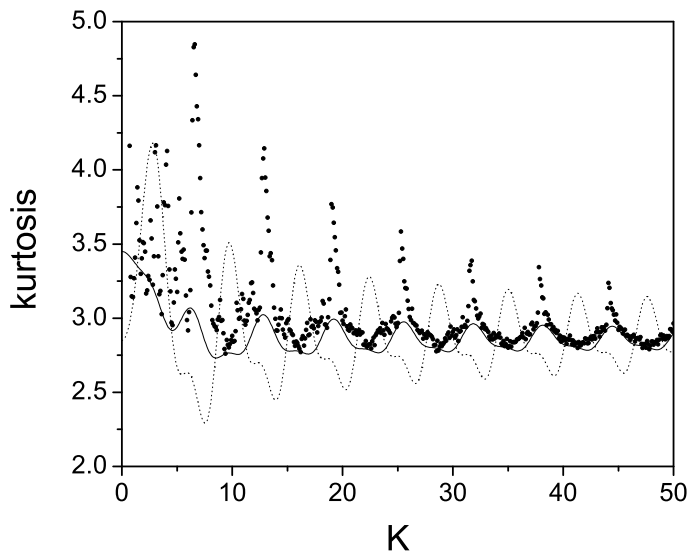}}
\end{center}
\caption{Kurtosis as a function of $K$ for the standard map for $c=1$ and $n=10$. The dots
 correspond to   numerically computed values and the solid line to the theoretical result in the chaotic regime, up to terms of order $\mathcal{O}(K^{-1})$. For each value of $K$, $10^{5}$ random initial conditions are considered in the numerical simulation. This plot shows good agreement with numerical calculations although the accelerator modes give rise to spikes in the figure. The dashed line corresponds to the Kurtosis calculated from the
 results of   Tabet {\it et. al} presented in \cite{Tabet}.  See   Section \ref{comparison} for further
 details.
 As one can see, the present results give rise
 to a superior description. }
\label{fig1}
\end{figure}

The standard map corresponds to the choice $c\alpha(I)=I$ and $f(\theta)=\sin\theta$ in
(\ref{map}).
The memory functions (\ref{psi0})-(\ref{psi2}) are rather simple for the
 case of   linear rotation number since one has $\mathcal{G}_{\lambda}(1,x)=\delta_{\lambda,x}$.
For the standard map,
 $\mathcal{J}_{m}(x)$ is the Bessel function of the first kind $J_{m}(x)$, $D_{ql}=K^{2}/4$ and $\sigma_{m,m'}=(\pm\delta_{m,\pm1})(\pm\delta_{m',\pm1})$. The final expressions for $D$ and $B$ in the chaotic regime for the standard map  are given by
\begin{eqnarray}
\label{dsm}
\frac{D}{D_{ql}}&=&1-2J_{2}(K)+\mathcal{O}(K^{-1}), \\
\frac{B}{D_{ql}^{2}}&=&-\frac{1}{4}+J_{0}(K)+2J_{2}(K)+J_{4}(K)+\frac{1}{2}J_{4}(2K)+ \mathcal{O}(K^{-1}).\label{bsm}
\end{eqnarray}
The kurtosis (\ref{kurt}) can be evaluated  straightforwardly. Fig. \ref{fig1}
 depicts the calculated kurtosis and the results of numerical simulations for the standard map.
 We notice, in particular, the presence of accelerator modes.

 \subsection{The sawtooth map}

\begin{figure}[ht]
 \begin{center}
\resizebox{0.65\linewidth}{!}{\includegraphics*{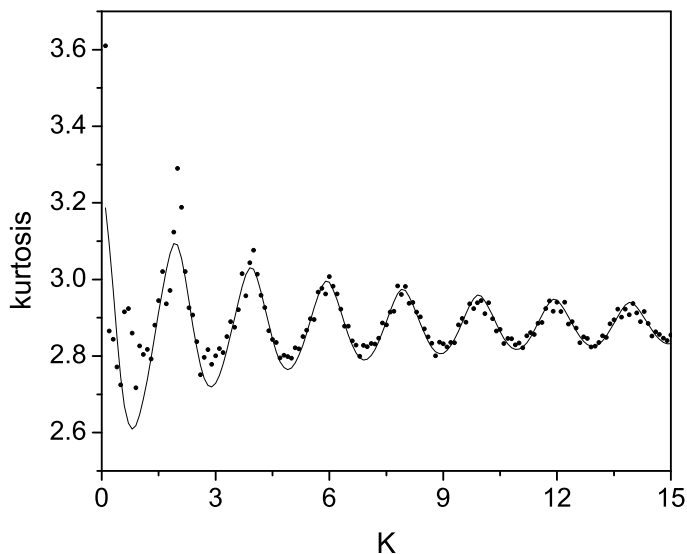}}
\end{center}
\caption{Kurtosis as a function of $K$ for the sawtooth map for $c=1$ and
$n=10$. For each value of $K$, $10^{5}$ random initial conditions are considered
in the numerical simulation. This plot shows excellent agreement between  the
theory and the numerical simulations.}
\label{fig2}
\end{figure}

 The sawtooth map corresponds to $c\alpha(I)=I$ and $f(\theta)=\theta$.
 As for the standard map, $\mathcal{G}_{\lambda}(1,x)=\delta_{\lambda,x}$.
 On the other hand, in this case,
  $\mathcal{J}_{m}(x)=\frac{\sin[\pi(m-x)]}{\pi(m-x)}$, $D_{ql}=K^{2}\pi^{2}/6$, and $\sigma_{m,m'}=\frac{6}{\pi^{2}}\frac{(-1)^{m-m'}}{mm'}$, leading to
\begin{eqnarray}
\label{st1}
\frac{D}{D_{ql}}&=&1-\frac{1}{6}S_{3}(K) , \\
\label{st2}
\frac{B}{D_{ql}^{2}}&=&-\frac{1}{5}-\frac{1}{3}S_{3}(K)+\frac{3+2K}{2+K}C_{4}(K) +\frac{9+10K+3K^{2}}{(2+K)^{2}}S_{5}(K) ,
\end{eqnarray}
both up to $\mathcal{O}(K^{-2})$ order,
where the function $S_{j}(K)$ and $C_{j}(K)$ are given by
\begin{eqnarray}
S_{2j+1}(K)&\equiv&\sum_{m=1}^{\infty}\frac{72\sin(\pi mK)}{(K+2)(\pi m)^{2j+1}}\,,\\ C_{2j}(K)&\equiv&\sum_{m=1}^{\infty}\frac{72\cos(\pi mK)}{(K+2)(\pi m)^{2j}}\,,
\end{eqnarray}
and $S_{2j}(K)=C_{2j+1}(K)=0$.
 Fig. \ref{fig2} presents the comparison of the   kurtosis calculated from (\ref{st1}) and
 (\ref{st2})
  with the numerical
 simulations for the sawtooth map.
 As one can see, the absence of accelerator modes contributes with the excellent agreement of the
 numerical simulation with the theoretical predictions.

 A pertinent comment here is that, as one can see from Figures \ref{fig1} and \ref{fig2}, for a fixed
$n$, the limit for the kurtosis $\kappa$ is not $3$ as $K\rightarrow\infty$. From the asymptotic values of
(\ref{dsm})-(\ref{bsm}) and (\ref{st1})-(\ref{st2}), it is easy to show that, for $n=10$, the kurtosis in the limit $K\rightarrow\infty$ tend to the values $57/20$ and $72/25$, respectively, for the standard and sawtooth maps.
Similar results hold also for other maps.
The conclusion that,
for fixed $n$, the limit of high stochasticity  is not enough to assure a Gaussian regime is interesting
and certainly would deserve a deeper analysis.

\subsection{The tangent map}

 We call the tangent map the choice of  $f(\theta)=\theta$ and $\alpha(I)=\tan(I/2)$ in (\ref{map}).
 It is our first example of a map with a non-linear rotation number.
The functions  $\mathcal{J}_{m}(x)$ are the same ones from the sawtooth map.
The functions  $\mathcal{G}_{\ell}(1,x)$ can be calculated by complex residues
from their definition (\ref{Gfunction}).
Introducing the variable $s=\tan(\theta/2)$ and taking into account the identity
$i\arctan s$ = ${\rm arctanh\,} is$, we have
\begin{equation}
\label{complexG}
\mathcal{G}_{\ell}(1,x) = \frac{1}{\pi}\int_{-\infty}^{\infty} \frac{(1-is)^{\ell -1}e^{ixs}}{(1+is)^{\ell+1}}
ds.
\end{equation}
For $\ell=0$, (\ref{complexG}) has two single poles on the complex plane located at $z=\pm i$. For positive
$x$ and   negative $x$, one  closes the integration path of (\ref{complexG}), respectively,
by the positive $\Im(z)$ and negative $\Im(z)$ semiplanes, giving simply
\begin{equation}
\mathcal{G}_{0}(1,x) = e^{-|x|}.
\end{equation}
Let us consider now $\ell >0$. Notice that $\mathcal{G}_{\ell}(1,x)$ can be evaluated for negative
$\ell$ by observing that $\mathcal{G}_{-\ell}(1,x) = \mathcal{G}_{\ell}(1,-x)$. The integral
(\ref{complexG}) has a unique pole of order $\ell+1$ at $z=i$ for $\ell>0$. In such a case, for
negative $x$ we can close the integration path in the negative $\Im(z)$ semiplane and
conclude that $\mathcal{G}_{\ell}(1,x)=0$ for $\ell>0$ and $x<0$. For positive $x$ we close
the integration path in the positive $\Im(z)$ semiplane and obtain
\begin{equation}
\label{gell}
\mathcal{G}_{l}(1,x)=
2\frac{(-1)^\ell}{\ell!}\left[\frac{d^\ell}{ds^\ell}\left((s+1)^{\ell-1}e^{-xs}\right)\right]_{s=1},
\end{equation}
for $\ell>0$ and $x\ge 0$. As   illustrative examples, the very first functions (\ref{gell}) are given by
\begin{eqnarray}
\mathcal{G}_{1}(1,x) &=& 2xe^{-x}, \\
\mathcal{G}_{2}(1,x) &=& 2x(x-1)e^{-x}, \\
\mathcal{G}_{3}(1,x) &=& \frac{2}{3}x(2x^2-6x+3)e^{-x},
\end{eqnarray}
for $x\ge 0$.

\begin{figure}[ht]
\begin{center}
\resizebox{0.65\linewidth}{!}{\includegraphics*{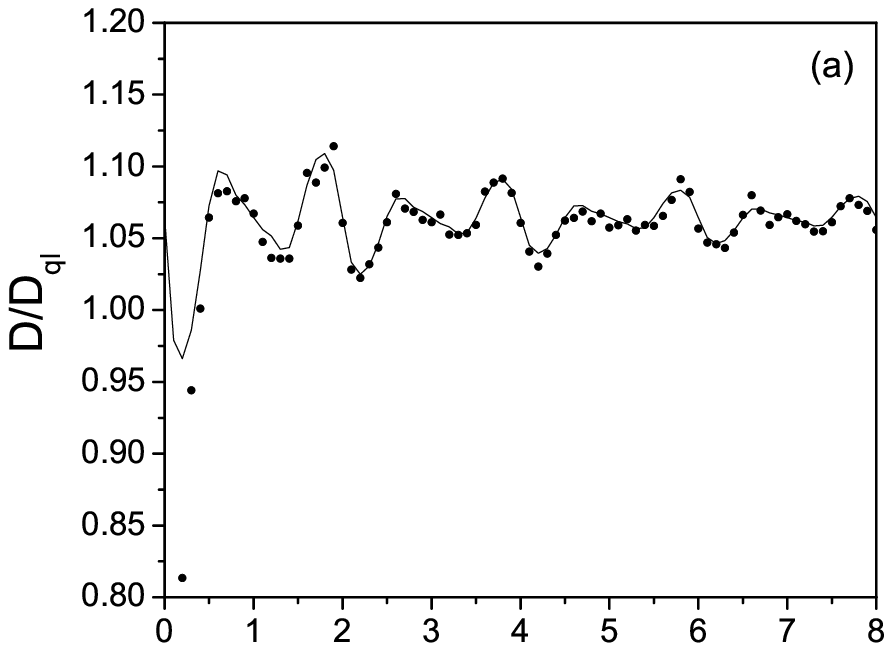}}
\resizebox{0.65\linewidth}{!}{\includegraphics*{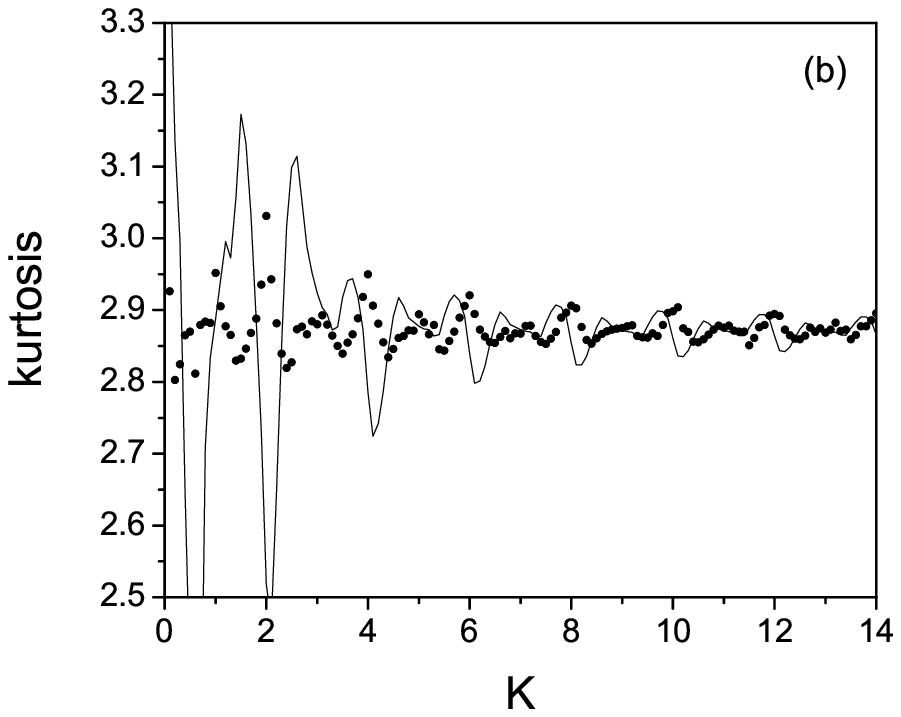}}
\end{center}
\caption{Diffusion (a) and Kurtosis (b)
as a function of $K$ for the tangent map with $c=3$ and $n=10$. The dots corresponds to
 the numerically computed values and the solid line to the theoretical result in the chaotic regime.
 For each value of $K$, $10^{5}$ random initial conditions are
  considered in the numerical simulations. Both plots also exhibit good agreement between the predictions and the numerical
  simulations.}
\label{fig3}
\end{figure}

The expressions for $D$ and $B$ are obtained from  (\ref{diffusionformula}) and
(\ref{2trunc}),
respectively. Fig. \ref{fig3} depicts the comparison between
the calculated diffusion and kurtosis for the tangent map and  the numerical simulations.

\subsection{The cubic map}
\label{cubicmap}
We call the cubic map the choice $\alpha(I)=I^3/3$ and $f(\theta)=\sin\theta$ in (\ref{map}).
It is our example of map with non-periodic non-linear rotation number.
Due to the non-periodicity,  some parts of our formalism shall be
modified.  First,
 we   rewrite the function  (\ref{Gfunction})  in the following way
\begin{equation}
\mathcal{G}_{\lambda}(r,x)=\frac{1}{2\pi r}\int_{-\pi r}^{\pi r}du\,e^{-i(x\alpha(u)-\lambda r^{-1}u)}.
\end{equation}
The non-periodic rotation number corresponds to the
 limit $r\rightarrow\infty$. In this case, we apply the following overall replacements in our formulas
\begin{eqnarray}
&  \quad \quad r^{-1}\lambda\longrightarrow s,&\\
&\quad r^{-j}\!\!\!\!\!\sum_{\left\{\lambda_{1}, \ldots, \lambda_{j}\right\}}\!\!\!\!\! \longrightarrow \int ds_{1}\times\ldots\times ds_{j}\,,&\\
&  r\,\mathcal{G}_{\lambda}(r,x)\longrightarrow\mathcal{G}(s,x),&
\end{eqnarray}
where the function $\mathcal{G}(s,x)$ is now given by
\begin{equation}
\mathcal{G}(s,x)=\frac{1}{2\pi}\int dI\,e^{-i(x\alpha(I)-sI)}.
\end{equation}
Performing analogous calculations to the periodic case, we obtain   the new set of memory functions
\numparts
\begin{equation}
\label{psi0np}
\Psi_{0}(q)=\mathcal{J}_{0}(-Kq)\,,
\end{equation}
\begin{equation}
\label{psi1np}
\Psi_{1}(q)=\sum_{m}\mathcal{J}_{-m}(-Kq)\mathcal{J}_{m}(-Kq)\,\mathcal{G}(0,mc)\,,
\end{equation}
\begin{eqnarray}
\label{psijnp}
\Psi_{j\geq2}(q)&=&\sum_{\{m\}}\mathcal{J}_{-m_{1}}(-Kq)\,\mathcal{J}_{m_{j}}(-Kq)\,\int_{\mathcal{S}^{\dag}}d\vec{s}\,\mathcal{G}(s_{1},m_{1}c)
\times \nonumber \\
&&\times
\prod_{i=2}^{j}\mathcal{G}(s_{i},m_{i}c)\,\mathcal{J}_{m_{i-1}-m_{i}}\left[-K\left(q+\sum^{i-1}_{k=1}s_{k}\right)\right],
\end{eqnarray}
\endnumparts
where $d\vec{s}=ds_{1}\times\dots\times ds_{j}$, with the domain of integration   defined by ${S}^{\dag}=\left\{(s_{1}, \ldots,s_{j}):\sum_{i=1}^{j}s_{i}=0\right\}$. Finally, the diffusion coefficient will be given by
\begin{eqnarray}
\label{diffusionformulanp}
\frac{D}{D_{ql}}&=&1+2\sum_{m=1}^{\infty}\sigma_{m,m}\,\mbox{Re}[\mathcal{G}(0,mc)]+\sum^{\infty}_{j=2}\sum_{\{m\}}\,\sigma_{m_{1},m_{j}}\,\int_{\mathcal{S}^{\dag}}\,d\vec{s}\,\mathcal{G}(s_{1},m_{1}c)\times\nonumber \\ && \times \prod_{i=2}^{j}\mathcal{G}(s_{i},m_{i}c)\,\mathcal{J}_{m_{i-1}-m_{i}}\left(-K\sum^{i-1}_{k=1}s_{k}\right).\nonumber\\
\end{eqnarray}
Note that, for non-periodic  linear rotation numbers, we have $\mathcal{G}(s,x)=\delta(s-x)$, and the diffusion formula (\ref{diffusionformulanp})   coincides with the periodic  linear rotation number version of (\ref{diffusionformula}).

For the cubic map, the function $\mathcal{G}(s,x)$
can be calculated by means of
 Airy functions\cite{Abramowitz}
\begin{equation}
\label{GAiry}
\mathcal{G}(s,x)=x^{-1/3}\mbox{Airy}(-x^{-1/3}\,s).
\end{equation}
The power dependence $x^{-1/3}$ in (\ref{GAiry}) may create a false impression of divergence of the series (\ref{diffusionformulanp}) for the cases where $c^{1/3}\ll 1$. In order to avoid this problem, one
can define
\begin{equation}
\label{c1c2}
c\equiv c_{1}c_{2}, \qquad c_{2}^{1/3}I\equiv x,
\end{equation}
and rewrite the cubic map as
\begin{eqnarray}
\label{ssx}
\begin{array}{l} x_{n+1} = x_{n}+Kc_{2}^{1/3}\mbox{sen}\,\theta_{n}\,,\\
                 \theta_{n+1} = \theta_{n}+\,c_{1}\,x_{n+1}^{3}/3\qquad \mbox{mod}\, 2\pi.
\end{array}
\end{eqnarray}
From (\ref{c1c2}), it follows that $(\Delta x)^{2}=c_{2}^{2/3}(\Delta I)^{2}$ and $D_{ql}^{(x)}=c_{2}^{2/3}D_{ql}^{(I)}$. Hence, the rate $D/D_{ql}$ for the cubic map is invariant under the
rescaling
\begin{equation}
\label{escal}
\frac{D}{D_{ql}}\,(c,K)\equiv\frac{D}{D_{ql}}\,(c_{1}c_{2},K)=\frac{D}{D_{ql}}\,(c_{1},Kc_{2}^{1/3}).
\end{equation}
Assuming that
\begin{equation}
K\gg c^{-1/3},
\label{rshc}
\end{equation}
the rescaling
\begin{equation}
c_{1} = K^{3}c,\qquad c_{2}= K^{-3}
\end{equation}
prevents any potential problem of divergence for maps with small $c$.

\begin{figure}[ht]
\begin{center}
\resizebox{0.65\linewidth}{!}{\includegraphics*{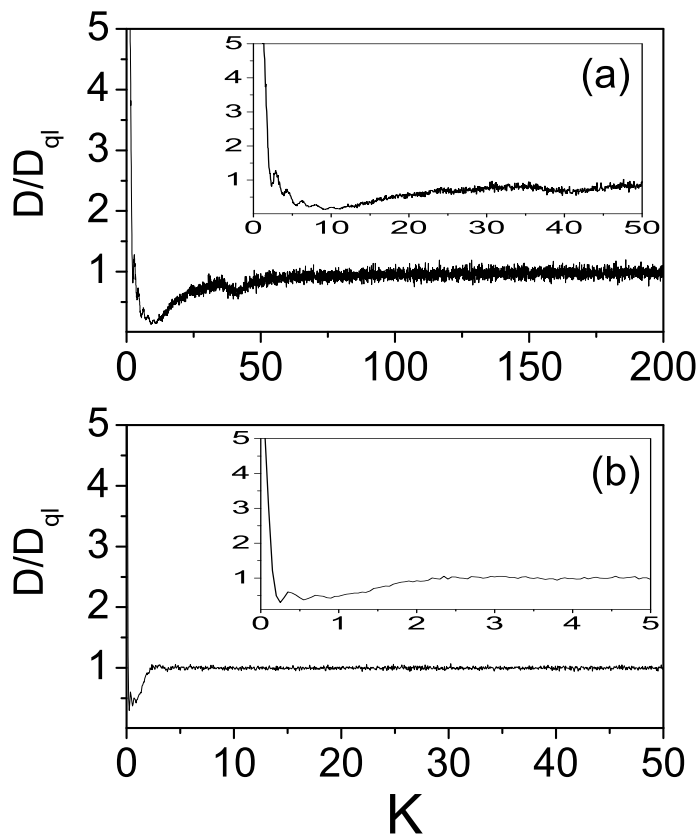}}
\end{center}
\caption{
Numerically calculated diffusion for the cubic map with $c=10^{-4}$ (a) and $c=10^{-1}$ (b).
For each value of $K$, $10^{5}$ random initial conditions are
  considered in the numerical simulations.
According to our results (see Section \ref{cubicmap}),   quasilinear behavior is expected to occur for
$K\gg c^{-1/3}$. From the figures, quasilinear diffusion clearly takes place for
$K > 21$ and $K>2$, respectively, in accordance to our theoretical predictions.}
 \label{fig4}
\end{figure}

For small values of $K$, the memory functions $\Psi_{j\geq2}(q)$ give rise to
high oscillatory combinations of Airy and Bessel functions, with integrals that
are very difficult to estimate. On the other hand, the high stochastic condition (\ref{rshc}) implies that $\mbox{Airy}(c_{1}^{-1/3}s)\approx\mbox{Airy}(0)$ even for sufficiently high values of $|s|$ for which   Bessel functions already decay as $J_{m}(s) \sim e^{is}|s|^{-1/2}$. Thus, in this regime, the rate $D/D_{ql}$ for the cubic map can be estimated by
\begin{equation}
\left|\frac{D}{D_{ql}}-1\right|\leq\sum^{\infty}_{j=1}\frac{b_{j}}{(Kc^{1/3})^{j}},
\end{equation}
where
\begin{eqnarray}
b_{1}&=&2\mbox{Airy}(0),\\
b_{j\geq2}&=&\mbox{Airy}^{j}(0)\sum_{\{m\}}\frac{(\pm\delta_{m_{1},\pm1})(\pm\delta_{m_{j},\pm1})}{(m_{1}\times\ldots\times m_{j})^{1/3}}\,\times \nonumber \\
&&\times
\int\,ds_{1}\times\ldots\times ds_{j-1}\prod_{i=2}^{j}\,J_{m_{i-1}-m_{i}}\left(-\sum^{i-1}_{k=1}s_{k}\right)\nonumber\\
&=&\frac{1}{2}[2\mbox{Airy}(0)]^{j}\sum_{\{m_{i}=2l_{i}+1\}}\frac{(\pm\delta_{m_{1},\pm1})(\pm\delta_{m_{j},\pm1})}{(m_{1}\times\ldots\times m_{j})^{1/3}} =
8\mbox{Airy}^{2}(0)\,\delta_{j,2},
\end{eqnarray}
leading, as expected, to
$\lim_{K\rightarrow\infty}D=D_{ql}$. Indeed,
the quasilinear regime  for the diffusion is rapidly attained  for high values of $K$,
without    oscillations, see Fig. \ref{fig4}.
Similar results hold for any rotation number  of the type $\alpha(I)\propto I^{p}$, for $p>1$.
 In such cases,
 $\mathcal{G}(x,0)\propto x^{-1/p}$ and (\ref{rshc}) shall be replaced by $K\gg c^{-1/p}$.

\begin{figure}[ht]
\begin{center}
\resizebox{0.65\linewidth}{!}{\includegraphics*{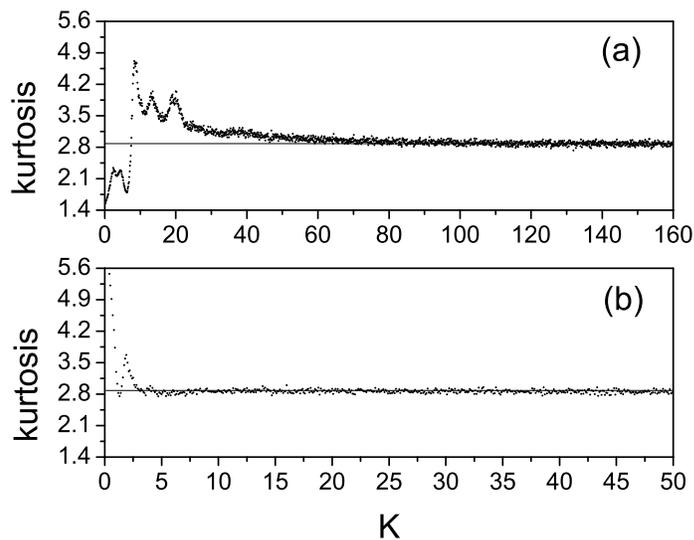}}
\end{center}
\caption{
Numerically calculated kurtosis for the cubic map with $c=10^{-4}$ (a) and $c=10^{-1}$ (b).
For each value of $K$, $10^{5}$ random initial conditions are
  considered in the numerical simulations.
As the diffusion coefficient (Fig. \ref{fig4}),   quaselinear behavior is expected to occur for
$K > 21$ and $K>2$, respectively.}
 \label{fig5}
\end{figure}

Regarding the kurtosis for the cubic map, similar arguments can be used to show that
\begin{equation}
\label{tttt}
\kappa \sim 3 - \frac{6}{5n},
\end{equation}
also without oscillations for $K\gg c^{-1/3}$. Since the diffusion is quasiliner
for such values of $K$, the memory functions $\Psi_j$ can be disregarded for
all $j\ge 1$. In this limit, from (\ref{2trunc}), we have simply
${B}/{D^2} \sim -{1}/{5}$,
and, hence, leading to (\ref{tttt}).
Fig. \ref{fig5} depicts the behavior of the kurtosis for the cubic map.

\subsection{Comparison with previous approaches}
\label{comparison}
It is instructive to compare our results with others previously
 obtained
  in the literature in the context of standard map. Tabet {\it et al}\cite{Tabet}, for instance,
   do not calculated the Burnett coefficient explicitly.  They used, instead, the Fourier path technique\cite{Lichtenberg,Rechester}
    to calculate the moments $\langle(\Delta I)^{4}\rangle_{n}$ and $\langle(\Delta I)^{2}\rangle_{n}$,
    from which
    we can evaluate the respective rate $B/D^{2}$ by means of the definitions (\ref{Ddef}) and (\ref{Bdef}). One has
\begin{equation}
\frac{B}{D^2}=-\frac{1}{4}+3J_{2}(K)+2J_{4}(2K)+\mathcal{O}(K^{-1}).
\label{TSDR}
\end{equation}
 On the other hand, Balescu\cite{Balescu2} utilizes a related, but somewhat different approach from ours to calculate   kurtosis for the standard map. He also does not calculate  explicitly the kurtosis nor the Burnett coefficient, but he present a non-Markovian approach for the relevant amplitudes, for
  which the propagator $F_{n}(q)$ is given by
\begin{equation}
F_{n}\approx\Psi_{0}^{n}+(n-2)\Psi_{0}^{n-3}\Psi_{2}+\frac{1}{2}(n-5)(n-6)\Psi_{0}^{n-6}\Psi_{2}^{2}
\label{Bpropag}
\end{equation}
for $n\geq6$, where
  memory functions $\Psi_{j\geq3}$   contributing with terms of order $\mathcal{O}(K^{-1})$ are
  disregarded. The corresponding Burnett coefficient can be calculated by means of (\ref{gammaF}) and (\ref{Bdef}), leading exactly to the same result we have gotten  here
 \begin{equation}
\frac{B}{D^{2}}=-\frac{1}{4}+J_{0}(K)+J_{2}(K)+J_{4}(K)+\frac{1}{2}J_{4}(2K)+\mathcal{O}(K^{-1}),
\label{BV}
\end{equation}
compare with (\ref{bsm}) and (\ref{dsm}).
 Although the equations (\ref{TSDR}) and (\ref{BV}) have the same asymptotic value, they are enough different even in the high stochastic regime. However, it becomes difficult to note differences in the kurtosis calculated in both cases for large values of $n$ (Tabet {\it et al} used $n=50$, for instance). For relatively smaller values, the equation (\ref{BV}) gives a much
 better agreement with numerical simulations than (\ref{TSDR}), as one can see in the Fig.\ref{fig1}.

\section{Gaussian  characteristic time scale}

Equation (\ref{kurt}) suggests the existence of a characteristic time: the { Gaussian time scale} $n_{G}$, defined by
\begin{equation}
n_{G}\equiv\max_{K}\left\lceil6(|B|/D^{2})\right\rceil,
\label{ng}
\end{equation}
where $\left\lceil x\right\rceil=\inf\{n\in Z|x\leq n\}$ is the {  ceiling function}. For
 $n \gg n_{G}$, the transport process is typically Gaussian, up to order $\mathcal{O}(q^{6})$.
  In fact, for $n \gg n_{G}$, the expansion of the propagator $\exp[n\gamma(q)]$ gives the well known Gaussian density
\begin{eqnarray}
\exp[n\gamma(q)]&=&1-Dnq^{2}+\frac{1}{2}Dn^{2}\frac{\kappa}{3}\,q^{4}+\mathcal{O}(q^{6})\nonumber\\
&\approx&\exp(-Dnq^{2})+\mathcal{O}(q^{6}),
\end{eqnarray}
where $\kappa$ is given by equation (\ref{kurt}). The Gaussian time scale $n_{G}$ has also a second and no
lesser  important interpretation: in the regime $n\gg n_{G}$, the time evolution of the relevant amplitudes becomes Markovian. This is easy to realize since, for $n\gg n_{G}$, the higher order corrections can be neglected and the LPR resonance (\ref{gammaq6approx}), which gives the exact expression for the normal diffusion
coefficient\cite{Venegeroles}, can be taken as the approximate propagator.
Hence,  for $n\gg n_{G}$,  equation (\ref{anq}) holds perfectly and, furthermore, becomes the following
Markovian master equation
\begin{equation}
a_{n}(q)\approx \left(\sum_{j=0}^{\infty}\Psi_{j}(q)\right) a_{n-1}(q).
\label{Markovian}
\end{equation}
The Gaussian time scale $n_{G}$ is sharper than some characteristic times   obtained  previously
in the literature, as, for
instance, Balescu's {  memory time} $n_{M}$ introduced  in \cite{Balescu2}.
Balescu   obtained
 the propagator (\ref{Bpropag})  by means of the general non-Markovian  Bandtlow and Coveney
 master equation\cite{Bandtlow}. The convolution of the master equation is truncated at  the {  memory time} $n_{M}$, leading to
\begin{equation}
\label{NM}
a_{n+1}(q)\approx\sum_{j=0}^{n_{c}}\Psi_{j}(q)\,a_{n-j}(q),
\end{equation}
where
\begin{equation}
n_{c}=\left\{ \begin{array}{ll}
      n\qquad\,\,\,\,\,\mbox{for}\,\,\,n\leq n_{M},\\
      n_{M}\qquad\mbox{for}\,\,\,n>n_{M}.
                 \end{array}
                 \right.\nonumber
\end{equation}
Notice that, for $n\gg n_{M}$, the equation (\ref{NM}) becomes the Markovian equation (\ref{Markovian}). By means of some numerical experiments with the decay of memory functions  for the standard map, Balescu concludes that $n_{M}=4$ and, in such a case, obtained equation (\ref{Bpropag}).

\begin{figure}[ht]
\begin{center}
\resizebox{0.65\linewidth}{!}{\includegraphics*{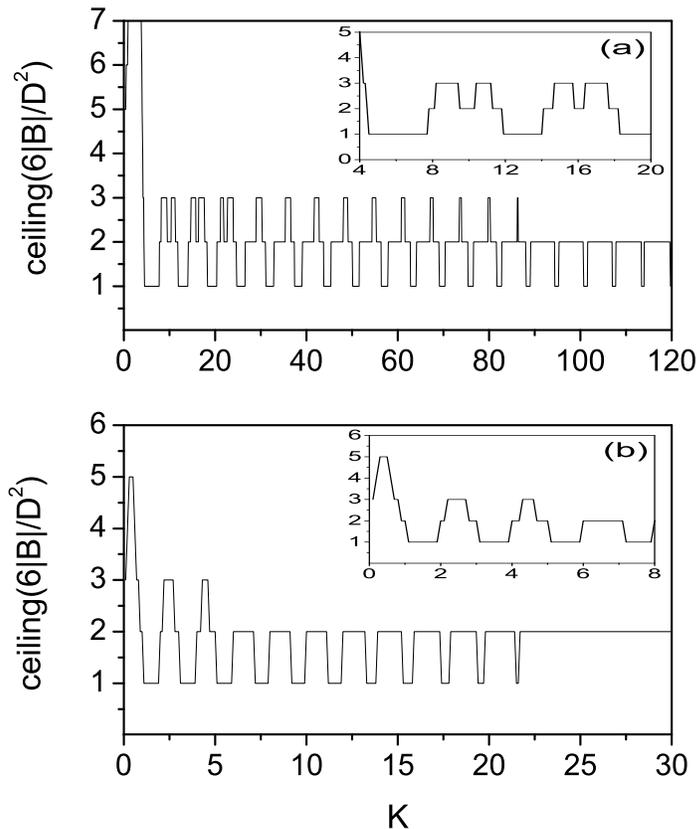}}
\end{center}
\caption{Gaussian time scale as a function of $K$ for the standard map (a) and
for the sawtooth map (b), both with $c=1$ and $n=10$.
For the standard map, we have $n_{G}=3$ (or $n_{M}=2$) for $K>4$ (for the
sawtooth map, for $K>2$),
in excellent agreement with our expectations. For larges $K$, $n_{M}$ falls to 1 indicating the absence of correlations of order $K^{-1/2}$ for large $n$, for both cases.}
\label{fig6}
\end{figure}

Evidently, we expect that the two time scales to be related by
$n_{G}=n_{M}+1$.
The advantage of   $n_{G}$ is that it can be calculated judiciously by means of (\ref{ng}) for a generic
 class of systems like (\ref{map}). Moreover, typically, $n_{G}$ is sharper than
 Balescu's memory time.
 In Fig. \ref{fig6}
we can observe the behavior of the rate $\left\lceil6|B|/D^{2}\right\rceil$ as a function of $K$ for the standard map and for the sawtooth map.
The theoretical prediction given by (\ref{ng}) is confirmed for $K>4$ for the
standard map ($K>2$ for the sawtooth map). However, it is important also to point out that the Gaussian time
 (\ref{ng}) evidently fails in the description of the weak-stochasticity regime, since the effects of the KAM surfaces and the stable islands become increasingly important as $K$ decreases, requiring not only the calculation of further memory functions but the inclusion of the source term involving initial conditions (see \cite{Bandtlow}).
 For instance, let us
  consider the equation (\ref{NM}) for the general linear rotation number
 case by applying the Balescu approach for $n_{M}=2$ (see Appendix). New calculations of the Burnett coefficient by means of the propagator (\ref{prapp}) gives
\begin{equation}
\label{BVconf}
B\approx-\frac{5}{2}D^{2}+2DD_{ql}+\frac{1}{4!}\partial_{q}^{4}(\Psi_{0}+\Psi_{2})_{q=0},
\end{equation}
that is exactly the equation (\ref{2trunc}) for $\Psi_{1}=0$! Hence, we should expect in the chaotic regime
\begin{equation}
\label{ng3}
n_{G}=3\qquad \mbox{for}\,\,\,\,\,c\,\alpha(I) = I.
\end{equation}

\section{Summary and Discussion}

In this paper, we have performed a detailed analysis of the non-Gaussian aspects of the normal transport in Hamiltonian discrete systems. The general class of area-preserving maps represented by (\ref{map}) has been chosen because it comprises a large number of physical situations and  has the
 paradigmatic
 standard map as a particular case. The map (\ref{map}) was recently studied in \cite{Venegeroles}, where the basis for the study of the higher order transport coefficients, including nonhyperbolic systems, was initiated. The LPR resonances of the system (\ref{map}) were enhanced in its wavenumber dependence with corrections of order $q^{4}$, so that the corresponding fourth order Burnett coefficient could be evaluated.  Numerical simulations were performed for four particular cases of (\ref{map})  and excellent agreement with the theoretical predictions is obtained.

We have established also a Gaussian time scale $n_{G}$ given by equation (\ref{ng}). For $n\gg n_{G}$, the density function acquires a Gaussian countour and its time evolution is ruled by a Markovian master equation. We also show that $n_{G}$ is related with the memory time $n_{M}$ defined by Balescu \cite{Balescu2} by $n_{G}=n_{M}+1$. For maps such that $c\alpha(I)=I$ we conclude that $n_{G}=3$, a sharper result than
Balescu's memory time $n_{M}$.

\ack
The authors thanks E. Gu\'eron for enlightening discussions  and
G. Dalpian for providing computing resources.
This work was supported by funds of FAPESP, CNPq, and UFABC. A.S.  is grateful  to   Prof. V. Mukhanov
for the warm hospitality at the Ludwig-Maximilians-Universit\"at, Munich, where   part of this work
was carried out.

 \appendix

\section{Balescu's approach for $n_{M}=2$}

In this appendix we  show that the cuttof time $n_{M}$ in the Balescu equation (\ref{NM}) can be taken as $n_{M}=2$
in the chaotic regime. For every calculation, we retain   arbitrary powers of $\Psi_{0}$ and terms of order $\Psi_{2}$ and $\Psi_{2}^{2}$ (recalling that $\Psi_{1}=0$ for the linear rotation number case).    Equation (\ref{NM}) in the form $a_{n+1}(q)=F_{n}(q)\,a_{n}(q)$ where
\begin{equation}
F_{n+1}(q)=\sum_{j=0}^{n_{c}}\Psi_{j}(q)F_{n-j}(q), \qquad F_{0}(q)\equiv1,
\label{A1}
\end{equation}
with $n_{c}=n$ for $n\leq n_{M}$ and $n_{c}=n_{M}$ for $n>n_{M}$. The next six propagators $F_{n}$ are given by
\begin{eqnarray}
F_{1}&=&\Psi_{0},\nonumber\\
F_{2}&=&\Psi_{0}^{2},\nonumber\\
F_{3}&=&\Psi_{0}^{3}+\Psi_{2},\nonumber\\
F_{4}&=&\Psi_{0}^{4}+2\Psi_{0}\Psi_{2},\nonumber\\
F_{5}&=&\Psi_{0}^{5}+3\Psi_{0}^{2}\Psi_{2},\nonumber\\
F_{6}&=&\Psi_{0}^{6}+4\Psi_{0}^{3}\Psi_{2}+\Psi_{2}^{2}.\nonumber
\end{eqnarray}
Thus, we can write a general expression for $F_{n}$:
\begin{equation}
F_{n}=\Psi_{0}^{n}+x_{n}\Psi_{0}^{n-3}\Psi_{2}+y_{n}\Psi_{0}^{n-4}\Psi_{2}^{2},
\label{A2}
\end{equation}
with the following initial conditions (note that $\Psi_{0}^{k}\Psi_{2}^{2}\sim\Psi_{2}^{2}\sim q^{4}$ for finite $k$)
\begin{equation}
\label{A1c}
\left\{ \begin{array}{ll}
       ~~~~x_{1}=0,\qquad x_{n\geq2}=n-2,\\
       y_{n\leq5}=0,\qquad~~~~~ y_{6}=1.
                 \end{array}
                 \right.
\end{equation}
On the other hand, the equation (\ref{A1}) gives
\begin{equation}
F_{n}=\Psi_{0}F_{n-1}+\Psi_{2}F_{n-3},\qquad n\geq3.
\label{A3}
\end{equation}
Substituting the equation (\ref{A2}) into (\ref{A3}) and comparing the coefficients we obtain:
\begin{equation}
\label{A4c}
\left\{ \begin{array}{ll}
       x_{n} = x_{n-1}+1,\\
       y_{n} = y_{n-1}+x_{n-3}.
                 \end{array}
                 \right.
\end{equation}
Solving the system (\ref{A4c}) with the initial conditions (\ref{A1c}) we finally obtain the propagator $F_{n}$ for $n\geq6$:
\begin{equation}
\label{prapp}
F_{n}=\Psi_{0}^{n}+(n-2)\Psi_{0}^{n-3}\Psi_{2}+\frac{1}{2}(n-5)(n-4)\Psi_{0}^{n-4}\Psi_{2}^{2},
\end{equation}
which may be compared to (\ref{Bpropag}).

\end{document}